\documentstyle[12pt,aasms4]{article}
\received{1999 July 2}
\accepted{1999 December 21}

\slugcomment{Accepted for publication in A\&A, Main Journal }

\lefthead{Ikuta \& Arimoto}
\righthead{Self-enrichment in $\omega$ Cen}

\begin{document}

\title{Self-enrichment in $\omega$ Centauri
    }

\author{C. Ikuta and N. Arimoto}
\affil{Institute of Astronomy, School of Science, 
              University of Tokyo, \\
              Mitaka, Tokyo 181-8588,  \\
              JAPAN}

\begin{abstract}

The origin of abundance spreads observed in $\omega$ Centauri 
is studied in the context of the self-enrichment scenario. 
Five chemical evolution models are constructed and  
are compared with empirical metallicity distribution 
of $\omega$ Cen.  After a series of simulations, it is found that 
neither of closed-box, outflow, nor infall models 
can reproduce the empirical metallicity distribution of $\omega$ Cen, 
while a modified outflow model with a bimodal initial mass function (IMF) 
gives a metallicity distribution that fits closely 
to the empirical ones. 
In the modified outflow model, long-lived stars are assumed to form after  
the first explosion of type II supernovae (SNII) in a proto-cloud. 
The modified outflow model involves gas infall at the very first
chemical evolution. Thus we conclude that self-enrichment causes 
the abundance dispersion in $\omega$ Cen.  
A success of the outflow model with the bimodal IMF implies that 
low mass stars in a globular cluster (GC) 
should have formed in the gas already 
enriched by the first generation of SNII. 

This scenario, originally proposed by Cayrel (1986), 
can explain a lack of globular clusters with [Fe/H]$\la -2.2$ 
in the Milky Way Galaxy. 

\end{abstract}


\section{Introduction}

The first comprehensive study of a peculiar globular cluster 
$\omega$ Centauri, including photographic 
photometry and proper motion analysis 
for several thousands stars, was undertaken by Woolley (1966).  
The resulting colour-magnitude diagram was discussed by 
Dickens \& Woolley (1967), who showed a large colour width 
of the red giant branch (RGB) stars of $\omega$ Cen. 
Cannon \& Stobie (1973) confirmed later the spread of RGB colours and 
argued that it must result from a heavy element abundance dispersion. 
Freeman \& Rodgers (1975) revealed a diversity of the composition of 
observed RR Lyrae variables and suggested that it may come 
from metal enrichment of the gas during proto-cloud collapse. 
Butler, Dickens \& Epps (1978) 
observed nearly half of the cluster RR Lyrae variables and 
found an intrinsic range in [Fe/H] from $-2.0$ to $-1.1$. 

However, the origin of abundance scatter observed 
in $\omega$ Cen is still an open question. 
Abundance anomalies of C, N, O, Na, Mg, and Al 
are often observed in Galactic globular clusters (GCs), 
and are interpreted as being a mixing effect at the later stage 
of stellar evolution (e.g., Kraft 1994). Unlike other GCs, 
$\omega$ Cen and probably also M22 (Lehnert, Bell, \& Cohen 1991) show the 
abundance spreads~ of ~the~ iron-peak~ elements.~
Since~ no~ stellar \\ evolution 
theory predicts changes in the abundances of these elements,   
this is one of the interesting features of the cluster. \\
Omega Cen~ is~ the~ brightest~ (Harris \& Racine 1979), \\
potentially the most massive, and the dynamically youngest cluster 
(Trager, King, \& Djorgovski 1995) in the Milky Way Galaxy. 
Omega Cen is also one of 
the most flattened clusters (e.g. Freeman \& Norris 1981). 
Meylan \& Mayor (1986) suggested that the flattening results from  
rotation, because the rotational velocity tightly correlates with 
the ellipticity. 
Since $\omega$ Cen has lots of peculiarities 
and locates relatively close to the solar system 
($\sim 5.2$ kpc; Meylan 1987), photometric and spectroscopic 
observations have been carried out intensively. 
Despite these efforts, the process of chemical enrichment and 
the causes of abundance spreads are not yet well understood.  
Kraft (1979) suggested four possible explanations, 
and Smith (1987) discussed them again in his review paper: 
(i) the proto-cloud of $\omega$ Cen already had abundance inhomogeneity; 
(ii) stars in $\omega$ Cen were enriched 
when it passed by a Galactic disc; 
(iii) $\omega$ Cen was formed via a merger of less massive GCs; 
(iv) $\omega$ Cen was self-enriched. 

In this paper, focusing on the abundance scatters of elements, 
we investigate whether a self-enrichment scenario can explain 
observations of the chemical properties.  
We also cannot completely dismiss the first scenario. However,  
it is rather {\it ad hoc} to postulate a primordial cloud abundance 
inhomogeneity and it is still necessary to explain 
how the abundance dispersions 
were brought into a proto-cloud and why this phenomenon 
is conspicuous in $\omega$ Cen but not in other GCs. True, 
dust segregation may partly explain the abundance spreads 
in the proto-cloud. 
Bhatt (1988) suggested that the abundance scatter should be
enhanced by the dust segregation, since the dust collecting heavy elements 
fell into the proto-cloud centre by gravitational 
and viscous forces, and consequently the inner region attains higher  
metallicity than the outer.  
Equation\,(7) of Bhatt (1988) gives $0.1$\,Gyr as a time required to grow 
a metallicity difference up to $\sim 1$\,dex.  
On the other hand, stars of the first generation in the proto-cloud 
should be formed in a free-fall time scale,  
typically $\sim 0.01$ Gyr (e.g., Cayrel 1986).  
Lifetimes of massive stars are around a few Myr. 
This means that dust grains were disintegrated by UV radiation 
from newly formed stars before the abundance difference 
reaches to $\sim 1$ dex. 
Thus, the dust segregation hypothesis alone cannot explain 
the abundance scatters of $\omega$ Cen. 

The accretion scenario is not plausible either. 
Most GCs show little evidence for the scatter (Kraft 1979). 
This is also true for GCs which are on the 
orbit that closely passes by the Galactic disc (e.g. Dauphole et al. 1996), 
despite this they are most likely to be contaminated. 
Observed [$\alpha$/Fe] ratios are over-abundant (Norris \& Da Costa 1995), 
which implies that stars in $\omega$ Cen 
were not enriched by type Ia supernovae (SNIa) 
which synthesise a large amount of iron-peak elements. 
If stars in $\omega$ Cen were contaminated by the disc gas, 
[$\alpha$/Fe] should be close to solar, since  
the disc gas has nearly solar composition (Edvardsson et al. 1993). 
Thus, we put aside the second possibility. 

Third, the merging formation scenario is not applicable.   
Norris et al. (1997) favoured this scenario because, 
(i) the metallicity distribution (MD; 
Norris, Freeman, \& Mighell 1996, hereafter NFM96) is bimodal; 
(ii) N-body simulations by Makino, Akiyama \& Sugimoto (1991) 
predicted that merging of GCs result in a flattened system, 
and more significantly (iii) the metal-rich ([Ca/H] $> -1.2$) giants 
radially concentrate at the inner region  
than the metal-poor ([Ca/H] $\le -1.2$) ones and 
rotate differently. 
First, the bimodal MD claimed 
by NFM96 was not confirmed by 
Suntzeff \& Kraft (1996; hereafter SK96) 
who derived the  MDs for giants and subgiants, individually. 
If the merger creates a bimodal MD, the metallicity dispersions 
must already have been in both pre-merging clusters. 
Searle (1977) predicted that a merged cluster should have a bimodal MD. 
In his model, all the GCs have metallicity dispersions. 
Thus, Searle's model is not applicable for the Galactic GCs.  
Second, 
the N-body simulation performed by Makino et al. (1991) is for a binary GC.  
On the other hand, no binary GC has been discovered 
in the Milky Way Galaxy yet. 
It seems to be difficult for  
GCs to encounter randomly and to become a single system 
(Icke \& Alca\'\i no 1988) 
because orbital velocities ($v \ge 200$ km s$^{-1}$; Dauphole et al. 1996) 
of observed GCs 
are much larger than their internal velocity dispersions 
($\sigma \sim 10$km s$^{-1}$; Cudworth 1976). 
Third, the observed number of metal-rich stars 
is small and is affected by a sampling bias (NFM96; SK96). 
Thus it is not conclusive that
the metal-rich and the metal-poor stars have 
different positional and dynamical properties. 

Therefore, we suppose that the self-enrichment 
causes the abundance anomalies in $\omega$ Cen.
Cayrel (1986) investigated the reason why no population III GCs 
have been discovered yet 
and found that it is due to an enrichment by the first stars 
born in a cloud with the primordial abundance. 
Simulating the evolution of supernova remnants in a cloud, 
Morgan \& Lake (1989) concluded that every GC can be 
enriched at least by one SNII. 
Since $\omega$ Cen is a massive GC 
($\sim 3\,10^6 M_\odot$; Meylan \& Mayor 1986), 
it is likely that the enriched gas can be recycled 
and the metallicity scatters are produced. 
Based on accurate Str\"omgren photometry of $\omega$ Cen,
Hughes \& Wallerstein (1998) have recently shown that the most 
meta-poor stars are 14 Gyr old, the intermediate metallicity ones are 12 
Gyr old and the most metal-rich ones are 10 Gyr old. This might
give direct proof that $\omega$ Cen has produced stars of increasing
metallicity over a span of about 4 Gyrs, although the
resulting age span of course depends on stellar evolutionary tracks
adopted. Recently, NFM96 and SK96 obtained the metallicity distribution 
functions (MDF) of $\omega$ Cen. 
These studies show that the empirical MDF has a long tail 
extended toward higher metallicity and that 
an analytical MDF predicted by 
a simple model of chemical evolution fails to explain 
the observed MDF (NFM96; SK96).  

In this paper, we show that an outflow model can explain the observed MDF.  
We construct three models;  
outflow, infall, and modified outflow models. 
In the first two models, we assume that 
the mass range of newly formed stars is from $0.1M_\odot$ 
to $60M_\odot$ and time invariant. 
In the third model, which is the modified outflow model,  
the lower limit of newly formed stars s assumed to change 
after the first SNII explosions. Thus the IMF is not 
time invariant any more.  
The idea comes from Cayrel (1986). 
We find that 
the modified outflow model with the bimodal IMF provides the
metallicity distribution which gives an acceptable match to the data.  
If the numerical MDF should be more strictly compared with the 
observed one, in addition to the bimodal IMF and the outflow, the
gas infall is necessary at the very early stage 
of chemical evolution. 

The outline of the paper is as follows.
In section 2, we review the observed MDF. In section 3, 
the model prescriptions are given and the results are confronted with 
the observations, and in section 4, 
a chemical enrichment in GCs is discussed. We also discuss briefly 
spatial metallicity distribution and 
metallicity gradients observed in elliptical galaxies in section\,4. 
Our conclusions are summarised in section\,5. 

\section{The observed metallicity distribution}

For elements such as Fe and Ca,
long-lived stars should keep the original abundances
of the interstellar medium from which they formed.
The mixing in
stellar interior, such as convection and meridional circulation,
should have no effect on these elements, although light elements such as
C, N, and O are likely to change considerably.
Therefore, the metallicity distribution
is expected to provide information on chemical enrichment history and
is considered as a powerful clue to trace
the chemical evolution in $\omega$ Cen. In this section,
we compare the MDF derived by NFM96 and by SK96
and review the properties.

SK96 measured stellar metallicities of
bright giant stars  ($12.2 <$ V $< 12.8$; hereafter BG sample)
and fainter subgiant stars ($14.8 <$ V $<15.3$; hereafter SG sample)
using the near-infrared Ca\,{\small II} triplet lines at
$\lambda \lambda 8542-8662$\,\AA\,\,  
and
presented the [Fe/H] - MDF of each sample in the iron abundance scale.
On the other hand, NFM96 measured [Ca/H] based on
the near-infrared Ca\,{\small II} triplet
and the Ca\,{\small II} H and K lines and presented
the [Ca/H] - MDF. Since both authors measured equivalent widths W(Ca)
of the Ca lines and reduced equivalent widths W'(Ca)
which correlate with [Ca/H] better
than [Fe/H] (see Fig.\,7 in SK96),
we adopt [Ca/H] as a common scale.

Figure 1 gives a correlation between W'(Ca) and [Ca/H] for
the BG sample, where W'(Ca) is given as a function of $(V-V_{\rm HB})$,
which is a difference between the observed $V$ magnitude
of a star and that of the
horizontal branch (HB) of a cluster (SK96):
\begin{equation}
{\rm W'(Ca)}={\rm W(Ca)}+0.62(V-V_{\rm HB}).
	\label{eqn:wwbg}
\end{equation}
In Figure 1, [Ca/H] is taken from
Norris \& Da Costa (1995) who observed 40 stars in $\omega$ Cen
with high-dispersion spectroscopy.  
We derive the following formula for 
$-1.5 \le {\rm [Ca/H]} \le -0.3$ 
by a least-square fitting:
\begin{equation}
{\rm [Ca/H]}=-2.62+0.44{\rm W'(Ca)}.
	\label{eqn:w'}
\end{equation}
For stars out of this range,
we extrapolate equation\,(2). 
We do not consider 
that this dramatically changes the metallicity distribution 
at [Ca/H] $< -1.5$, 
since the number of such metal-poor stars is small, as we see below. 
The frequency distribution of [Ca/H]
at [Ca/H]$ > -0.5$ will be discussed in detail later.   
The following
equation (SK96) is used for the SG sample
instead of equation (\ref{eqn:wwbg}):
\begin{equation}
{\rm W'(Ca)}={\rm W(Ca)}+0.35(V-V_{\rm HB})-0.19,
	\label{eqn:wwsg}
\end{equation}
and calculate [Ca/H] by using equation (\ref{eqn:w'}).

Figure 2 shows the three MDFs, where the dashed line represents the sample
taken from NFM96, while the solid and the dotted lines
represent the BG and the SG samples of SK96, respectively.
Usually an observed MDF is presented in the form of a histogram.
However, we prefer to use a continuous distribution function
to avoid artificial distortion by a binning and to take into account
observational uncertainties properly (Searle \& Zinn 1978;
Laird et al. 1988).
The definition of such a continuous MDF $P(z)$ is
given by Laird et al. (1988):
\begin{equation}
P(z)=\frac{1}{N \sigma \sqrt{2 \pi}}
\sum_{i=1}^{N} \exp\left (-\frac{(z-z_i)^2}{2 \sigma^2} \right ),
	\label{eqn:mdf}
\end{equation}
where $\sigma$ gives a typical error in the observed values
$z_1$, $z_2$, ..., $z_N$ and is adopted $\sigma=0.05$ for the data
by NFM96 and $\sigma=0.07$ for these by SK96, respectively.

In Figure~2, we present the MDF for [Ca/H]$< -0.5$ following NFM96,
because sampling bias is significant at the higher abundance, 
a more detailed discussion is given in SK96.
By observing a magnitude limited sample,
data is suspected to be biased and metal-rich stars might be
under-represented,
because a fraction of stars brighter than
a given magnitude is a function of abundance.
In equations (\ref{eqn:wwbg}) and (\ref{eqn:w'}),
we assume that stars of the BG sample and those taken from NFM96
lie on the RGB rather than on the asymptotic giant branch (AGB).
Although the abundance distribution of AGB stars may
differ from that of the RGB stars because of
dependence of post-RGB evolution on mass and abundance,
inclusion of the AGB stars will not seriously distort
the observed MDF (NFM96).
Thus, it seems reasonable to consider that the MDF in Figure\,2
represents the global MDF of $\omega$ Cen.

The MDFs of SK96 and NFM96 samples are almost identical.
Both rise sharply at [Ca/H]$\simeq -1.7$
and reach a well-defined peak at [Ca/H]$\simeq -1.4$, and then gradually
decline. The MDF is characterised by a tail extended to higher abundance
which contains more stars than a
prediction of a simple chemical evolution model (NFM96; SK96).
There is a small bump at [Ca/H]$\simeq -0.9$
in the NFM96 sample, at [Ca/H]$\simeq -1$ and [Ca/H]$\simeq -0.6$ in
the BG and the SG samples of SK96, respectively.
We do not regard these bumps as significant, because of
the under-sampling effect mentioned before.
NFM96 inferred that the stars with [Ca/H]$\simeq -0.7$
are under-sampled by some 40\%
with respect to those having [Ca/H]$\simeq -1.7$.  \\

One might consider these second peaks as an effect of SNIa
which produce not only the iron-peak
elements but also calcium (Nomoto, Thielemann, \& Yokoi 1984).
However, [$\alpha$/Fe] of stars in $\omega$ Cen is super-solar
(Norris \& Da Costa 1995),  which means that the 
stars in $\omega$ Cen were formed in clouds which were not enriched by
SNIa.
In other words, duration of star formation must be shorter than
a typical lifetime ($\sim 1-2$\,Gyr) of SNIa progenitors, and thus
these bumps should not be related to SNIa.

We regard the well-defined peak, the sharp distribution,
and the metal-rich tail as
the key features of the empirical MDF of $\omega$ Cen.
A degree of the asymmetry of a distribution function 
can be described by a statistical value, skewness.
A positive value of skewness
signifies a distribution with an asymmetric tail extending out toward
larger measured value,
while a negative value signifies a distribution whose tail
extends out toward smaller value.
The definitions and more detailed descriptions
are given in Ikuta \& Arimoto (1999).
The skewness for the observed MDF of NFM96 is $0.73$,
which tells again that the MDF has a large asymmetric 
tail to the higher abundance.
We will use the skewness to constrain chemical evolution models.

\section{Theoretical metallicity distribution}

\subsection{Closed box and outflow models}

A chemical evolution model traces the abundance changes in the
interstellar matter (ISM) of a region, from which the stellar abundance
distributions are derived. A detailed derivation of the fundamental
equations can be found in Tinsley (1980).
We assume well-mixing of the interstellar matter and
relax the instantaneous recycling approximation
throughout this paper.
In this subsection,
we study a closed-box model and an outflow model following Hills (1980)
who pointed out that the GCs are slowly mass-losing systems.

Generally stellar birth rate is separated into two
independent functions. The birth rate of stars with mass
between $m$ and $m+dm$ is described as $C(t)\phi(m)dm$,
where $C(t)$ and $\phi(m)$ are the star formation rate (SFR) and
the IMF, respectively.
In this model, the IMF is assumed to be a time invariant function
with a power-law spectrum. Normalising to unity, we have
\begin{equation}
\phi(m)=\frac{(x-1)m_l^{x-1}}{1-(m_l/m_u)^{x-1}} m^{-x},
	\label{eqn:imf1}
\end{equation}
where the lower and the upper mass limits are
$m_l=0.1M_\odot$ and $m_u=60M_\odot$,
respectively.
The Salpeter IMF has $x=1.35$ in this definition.
The SFR is assumed to be proportional
to the gas fraction $f_{\rm g}(t)$:
\begin{eqnarray}
C(t) =\frac{M_{\rm T}}{\tau_{\rm sf}} f_{\rm g}(t) = \frac{M_{\rm
g}(t)}{\tau_{\rm sf}},
	\label{eqn:sfr}
\end{eqnarray}
with
\begin{equation}
f_{\rm g}(t)\equiv M_{{\rm g}}(t)/M_{\rm T},
\end{equation}
where $\tau_{\rm sf}$ and $M_{\rm T}$ are the time scale of star formation
and the initial total mass of $\omega$ Cen, respectively.
The gas mass $M_{\rm g}(t)$ changes through star formation and
gas outflow:
\begin{equation}
\frac{dM_{\rm g}(t)}{dt}=-C(t)+E(t)-B(t),
	\label{eqn:fg1}
\end{equation}
where $E(t)$ is the gas ejection rate from dying stars. 
$B(t)$ is the gas outflow rate (OFR) and is assumed to be given
by $M_{\rm g}(t)$ and the outflow time scale $\tau_{\rm of}$,
\begin{equation}
B(t)=\frac{M_{\rm T}}{\tau_{\rm of}}f_{\rm g}(t) 
=\frac{M_{\rm g}(t)}{\tau_{\rm of}}.
	\label{eqn:ofr}
\end{equation}
The equation governing the evolution of the abundance of $i$-th element
$Z_i(t)$ is given by:
\begin{equation}
\frac{d(Z_iM_{\rm g})}{dt}=-Z_i C(t)+E_{Z_i}(t)+Z_i(0)M_{\rm T},
	\label{eqn:zi}
\end{equation}
where $E_{Z_i}(t)$ is the total ejection rate of
processed and unprocessed $i$-th element and
$Z_i(0)$ is the initial abundance of $i$-th element in
a proto-cloud.
We calculate $E(t)$ and $E_{Z_i}(t)$ from the following integrals:
\begin{equation}
E(t)= \int\limits_{m_t}^{m_u}(1-w_m) C(t-t_m) \phi(m)dm,
\end{equation}
\begin{equation}
E_{Z_i}(t)\!=\!\!\!\int\limits_{m_t}^{m_u}\!\!\left[{(1\!-\!w_m)Z_i(t\!-\!t_m)\!+\!p_{im}}C(t\!-\!t_m)
\phi(m)\right]\!dm,
\end{equation}
where $t_m$ is the lifetime of a star with mass $m$. The lower limit
$m_t$ is the stellar mass with lifetime $t_m=t$.
Nucleosynthesis data $w_m$ and $p_{im}$ are the remnant mass fraction
and the mass fraction of processed and unprocessed
$i$-th element, both are taken from Thielemann, Nomoto, \& Hashimoto
(1996).
Although stellar metallicity could affect both $w_m$ and $p_{im}$,
this is still debated.
Thus we ignore these possible effects
and adopt the values calculated for the solar abundance.
We assume that the lower mass limit $m_{l,{\rm II}}$
for a progenitor of SNII is $10M_\odot$ (Tsujimoto et al. 1995).

There are two different points of view to explain the initial
metallicity of GCs:
(1) a proto-cloud is enriched by itself, (2) a proto-cloud is formed
from previously enriched gas.
Following Lee et al. (1995), we assume that a proto-cloud of
$\omega$ Cen was enriched and initially had a homogeneous composition.
Hence, the initial metallicity $Z_i(0)$ of the proto-cloud
is treated as a parameter.

Aiming to find a good model prescription,
we simulated $\sim 60000$ models,
since the parameters of chemical evolution such as
$\tau_{\rm SF}$, $\tau_{\rm OF}$, $x$, and duration of star formation
$T$ are unknown.
Fitting goodness is evaluated by a $\chi^2$-test with 
five degrees of freedom.
Model grids are the following:
\begin{eqnarray*}
	\begin{array}{lclrl}
\tau_{\rm sf} & = &0.001, \, 10^{-2.0+0.3i}~ (i=0,...,10), 
 3,\,5 ~~{\rm Gyr},\\  
\vspace{0.1cm}
\tau_{\rm of}& = & 0, \, \,0.001,\, 
\left[ 10^{-2.0+0.3i}~ (i=0,...,10) \right], 
 3,\, 5 ~~{\rm Gyr}, \\
\vspace{0.1cm}
x & = & 0.7 + 0.2i ~(i=1,...,7), \\
T & = & 10^{-2.0+i \Delta T}~(i=0,...,29)~~{\rm Gyr},\\
	\end{array}
\end{eqnarray*}
${\rm [Ca/H]}_0 = -\infty, -1.6,-1.5, -1.4$,\\
\vspace{0.05cm}\\
where $\Delta T=7.25 \cdot 10^{-2}$ Gyr, 
[Ca/H]$_0$ is the initial calcium abundance, 
and $\tau_{\rm of}=0$ corresponds to a closed-box model.

In Figure 3, the solid line represents a model which gives the
smallest $\chi^2=16.2$, and the dashed line represents
the empirical MDF by NFM96.
The model has $\tau_{\rm sf}=0.68$\,Gyr, $\tau_{\rm of}=1$\,Gyr,
$x=1.9$, and [Ca/H]$_0=-1.5$.
The best fit MDF is realized at $T=1.27$\,Gyr.
In despite of the good $\chi^2$ value,
it is obvious from eye judgement that the theoretical
MDF is inconsistent with the observations.~ 
There~ are~ fewer~ stars~ 
of [Ca/H] $ < -1.6$ and [Ca/H] $ > -0.8$,
while more stars of $-1.3 <$ [Ca/H] $< -1$.
The observations show that the lowest stellar metallicity
in $\omega$ Cen is around [Ca/H] $\simeq -1.7$.
This might imply that the mother cloud of long-lived stars 
in $\omega$ Cen
had already this metallicity. However,
too many stars with the initial abundance are formed when
we assume [Ca/H]$_0 = -1.7$ and such models are also
inconsistent with the observations. This resembles the G-dwarf
problem in the solar neighbourhood.

Although the theoretical MDF has a tail
slightly extended to the higher abundance, it is not long enough
compared with the empirical ones. 
The skewness quantitatively describes this. 
The skewness for the theoretical MDF (0.73)
is equal to 0.31, which is smaller than that of the
empirical MDF by at least a factor of two.
We find that a good model cannot be selected
based only on the $\chi^2$-test, since it is only an integrated
difference between a numerical and the observed MDF.
Therefore, we use the two statistics to determine good model
prescriptions.
A global fitting goodness is quantified by
the $\chi^2$-test, while
similarity of shape of the empirical and theoretical MDFs 
is measured by the skewness.

\subsection{Infall model}
To avoid the so-called G-dwarf problem,
we consider an infall model which is known to solve the G-dwarf problem
in the solar neighbourhood (Tinsley 1980).

Instead of equation (\ref{eqn:fg1}),
the evolution of the gas mass is given by
\begin{equation}
\frac{dM_{\rm g}(t)}{dt}=-C(t)+E(t)+A(t),
	\label{eqn:fg2}
\end{equation}
where $A(t)$ is a gas infall rate and is assumed
to be an exponentially decreasing function:
\begin{equation}
A(t)=\frac{M_{\rm T}}{\tau_{\rm in}}e^{-t/\tau_{\rm in}},
	\label{eqn:ifr}
\end{equation}
where $\tau_{\rm in}$ is an infall time scale.
The abundance evolution of $i$-th element is described as
\begin{equation}
\frac{d(Z_iM_{\rm g})}{dt}=-Z_iC(t)+E_{Z_i}(t)+Z_i'A(t),
	\label{eqn:zi2}
\end{equation}
where $Z_i'$ is an abundance of $i$-th element in the infalling gas.
Other model prescriptions are the same as those of the outflow model.
We perform more than 60000 simulations.
The model grids are the same as before except for
$\tau_{\rm in}$. We adopt $\tau_{\rm in}= 0.001,\,$
$\left[10^{-2.0+0.3i}~ (i=0,...,10) \right],
  \,$ $3,$\, $5 ~{\rm Gyr} $ instead of a sequence of $\tau_{\rm of}$, 
where we select a model by the $\chi^2$-test with five degrees of 
freedom. 

In Figure 3, the dotted line represents the MDF
of the best fit infall model $(\chi^2 = 62)$, 
whose $\tau_{\rm sf}$, $\tau_{\rm in}$, $T$, and $x$ are
0.25\,Gyr, 0.25\,Gyr, 0.034\,Gyr and 1.1, respectively.
This model initially had the primordial abundance.
We notice that the MDF of the infall model
has a tail extended to the lower abundance
and obviously fails to fit the observed MDF.
Calculating models assuming
$({\rm Ca/H})_0=0$ and/or a high infall rate,
we test and find that a tail of each MDF resulting from the infall
model always extends to the lower metallicity too much.
In conclusion, no good prescription is found in  
the outflow (including the closed-box) and the infall models.

\subsection{Models with a bimodal IMF}
The fitting goodness of the MDFs discussed in sections\,3.1\,\,and\,\,3.2
suggests that the outflow model is more or less favoured.
The essential problem of this model is that
too many stars with the initial abundance were formed. 
In the previous models, we have assumed
that the low and the high mass stars were born simultaneously. 
This assumption might be an over simplification 
when we study the first generation of stars in GCs.

Here, we consider another model proposed by Cayrel (1986) who
investigated the reason why no metal-free GCs have been observed.
He predicted that all GCs were self-enriched.
His basic argument is that the time scales 
of formation ($\sim 0.01$\,Gyr) and evolution ($\sim 0.001$\,Gyr)
of massive stars are shorter than
the corresponding time scale for low-mass stars.
In the free-fall collapse,
the density will be largest at the centre of a cloud, where
the first stars form.
The central stars are likely to be massive because of a shorter
time scale of fragmentation than that of low mass stars.
The first SNII explosions occur at the central core
toward which the gas is infalling.
The bulk of star formation with a full spectrum range
of stellar masses will occur in the shock between
the collapsing envelope and the SNII-driven wind.
In this view, low mass stars are expected to form
in the gas enriched by the first SNII explosions. 
This scenario can naturally explain the observed lower metal cutoff 
([Fe/H]$\simeq -2.2$) of GC systems
in the Milky Way Galaxy. 

Morgan \& Lake (1989) investigated a possibility of 
self-enrichment in a GC. By
simulating the evolution of supernovae remnants in a proto-cloud,
they examined whether the remnants can cool down
and become gravitationally bound.
They found that every GC should survive at least one supernova
explosion and can be enriched.
Thus we construct a model based on the scenario 
proposed by Cayrel (1986). 

\subsubsection{Bimodal IMF}
To take into account the delayed formation of low-mass stars
suggested by Cayrel (1986),
we assume that the lower mass limit $m_l$ of the IMF
is equal to $m_{l,{\rm II}}$ until the onset of the first SNII
explosion; that is,
star formation with a full mass spectrum occurs only
after the first SNII supernova explosion.
Equation (\ref{eqn:imf1}) is replaced by
\begin{equation}
\phi(m)=
\frac{(x-1)m_{l,{\rm II}}^{x-1}}{1-(m_{l,{\rm II}}/m_u)^{x-1}} m^{-x}
~~~(m_{l,{\rm II}} < m < m_u),
	\label{eqn:imf2}
\end{equation}
for $(t < t_{u,{\rm II}})$ and
\begin{equation}
\phi(m)=
\frac{(x-1)m_l^{x-1}}{1-(m_l/m_u)^{x-1}} m^{-x}
~~~(m_l < m <m_u),
	\label{eqn:imf3}
\end{equation}
for $(t \ge t_{u,{\rm II}})$,
where $t_{u,{\rm II}}$ is the lifetime of the most massive
SNII progenitor and is assumed to be equal to that of stars with $m_u$.
Since the total amount of metals which
are ejected by the first SNII is unknown, 
metallicity [Ca/H]$_0$ 
locked in the first generation of long-lived stars   
is taken as a parameter. 
Other model parameters 
are the same as those described in section\,3.1.
More than 60000 models are simulated and model grids are
the same as those of the outflow model.
As a result, we find two models 
(models A1 and A2) which give good $\chi^2$ with six degrees of freedom 
and skewness.  Their $\chi^2$ values, skewness, 
and parameters are given in Table~1.   

Models A1 and A2 have been selected 
ignoring whether the gas can be bound against
energy injection from SNII.
When the cumulative thermal energy $E_{\rm th}(t)$ from SNII
exceeds the gas binding energy $\Omega_{\rm g}(t)$,
the gas should be expelled rapidly
and hence the star formation must be terminated 
(Arimoto \& Yoshii 1987).
If the epoch ($T_{\rm w}$) of $E_{\rm th}(t)=\Omega_{\rm g}(t)$ 
comes before
a theoretical metallicity dispersion grows to $\sim 1$ dex,
the results of models A1 and A2 are unphysical.
Another possible case of gas ejection from a system is
that the entire volume of the system covered 
with supernovae remnants (SNRs) is 
expanding faster than the escape velocity at the centre
of the system (Ikeuchi, 1977).
This condition is not satisfied prior to the former one in any case,
since a SNR is rapidly cooled down due to high density
of the surrounding interstellar medium and can fill only a small area
(typically several pc in radius).
Thus we only discuss the evolution of
$E_{\rm th}(t)$ and $\Omega_{\rm g}(t)$.

The rotation curve of $\omega$ Cen presented 
by Meylan \& Mayor (1986)
implies that the dark matter distributes similarly 
to luminous matter and that
$\omega$ Cen does not have a massive diffused dark halo.
Therefore, the following equation (Saito 1979) is suitable
to describe $\Omega_{\rm g}(t)$:
\begin{equation}
\Omega_{\rm g}(t)=\Omega_{\rm T}f_{\rm g}(t) \left( 2-f_{\rm g}(t) \right),
	\label{eqn:bg}
\end{equation}
where $\Omega_{\rm T}$ is the total binding energy.
From analysing surface brightness distributions
and line-of-sight velocity dispersions,
Saito (1979) derived an empirical relation between $\Omega_{\rm T}$ and
$M_{\rm T}$ for spheroidals ranging from GCs to elliptical galaxies:
\begin{equation}
\Omega_{\rm T}=3.31\,10^{51} \left[\frac{M_{\rm
T}}{10^6M_\odot}\right]^{1.45} ~~{\rm erg}.
	\label{eqn:bt}
\end{equation}
Although equation (\ref{eqn:bt}) is based upon the present-day properties
of spheroidals, we assume that we can trace back the dynamical
state from information available today.
Using equation (\ref{eqn:bt}), the radius $R$ of a
viriarized spherical system is given as
\begin{equation}
R=13.1 \left[\frac{M_{\rm T}}{10^{6}~M_\odot} \right]^{0.55}~~\rm{pc}.
	\label{eqn:r}
\end{equation}
The virial theorem says that a system with no kinetic energy 
initially attains virial equilibrium by reducing a radius
to half of the initial value.  Assuming that the initial radius
is equal to $2R$ and that the matter in the system
distributes homogeneously, we can calculate the average density
$\rho(t)$:
\begin{eqnarray}
\rho(t) = \frac{3 M_{\rm g}(t)}{4 \pi (2R)^3} 
     = 7.27\,10^{-21} \left( \frac{M_{\rm g}(t)}{10^6 M_\odot}
\right)^{-0.65}
~~{\rm g\,cm}^{-3}
	\label{eqn:rho}
\end{eqnarray}

The cumulative thermal energy $E_{\rm th}(t)$ in the gas
is expressed as
\begin{eqnarray}
\begin{array}{l}
E_{\rm th}(t) = \left\{
\begin{array}{lr}
0  & (t < t_{u,{\rm II}})\\
\int^{m_u}_{{\rm max}(m_t, m_{l,{\rm II}})}dm  
\times \int\limits^{t-t_m}_{0}\!\frac{M_{\rm
T}}{m}\epsilon(t-t_m-t')C(t')\phi(m)dt &(t_{u,{\rm II}} \le t ). \\
\end{array} \right .
	\label{eqn:th}
\end{array}
\end{eqnarray}
The thermal energy content $\epsilon(t)$ inside a SNR is assumed
to evolve according to (Cox, 1972),
\begin{eqnarray}
\epsilon(t_{\rm SN})
&=& \!\!\!\!\left\{
\begin{array}{l}
7.2\,10^{50}\epsilon_0 ~~{\rm erg} \hspace{2cm} (0 \le t_{\rm SN} < t_{\rm c}(t)) \\
2.2\,10^{50}\epsilon_0(t_{\rm SN}/t_{\rm c}(t))^{-0.62} ~~{\rm erg} 
\hspace{0.4cm} (t_{\rm
c}(t) \le t_{\rm SN}),
\end{array} \right .
	\label{eqn:eth1}
\end{eqnarray}
where $\epsilon_0$ is an initial blast wave energy in units of $10^{51}$
erg,
$t_{\rm SN}$ is the time elapsed from onset of SNII explosion, and
$t_{\rm c}(t)$ is the complete cooling time of the SNR shell:
\begin{equation}
t_{\rm c}(t)=5.3\,10^{-5}\epsilon_0^{4/17}n(t)^{-9/17}~~{\rm Gyr}.
	\label{eqn:tc}
\end{equation}
We adopt $\epsilon_0=0.75$ (Cox 1972).
The surrounding gas density $n(t)$ in cm$^{-3}$ is assumed to
be equal to $\rho(t)/\mu m_{\rm H}$,
where $m_{\rm H}$ is the mass of a hydrogen atom and
$\rho(t)$ can be obtained from equation\,(\ref{eqn:rho}).
Neglecting masses of metals, we adopt
the mean molecular weight $\mu=1.3$, since we study a metal-poor
object.
Hereafter, we adopt $M_{\rm T}=3\,10^6\,M_\odot$ (Meylan \& Mayor
1986).
We check and confirm that the subsequent results do not change if
we adopt $M_{\rm T}=5\,10^6\,M_\odot$ (Meylan et al. 1995).

Figure\,4 gives the evolution of $E_{\rm th}/\Omega_{\rm g}$ of 
models A1 and A2.  
The epoch ($T_{\rm w}$) of $E_{\rm th}/\Omega_{\rm g}=1$  
is realized at $0.030$ and $0.042$~Gyr in models A1 and A2, 
respectively, which are described in Table\,1.  
On the other hand, 
the resulting durations of star formation of models A1 and A2 
are 0.03 and 0.06~Gyr, respectively. 
Comparing $T_{\rm w}$ with $T$ of each model, 
we reject model A2. Model A1 satisfies the criterion and 
the duration of star formation is  
also consistent with the argument in section\,2
that stars in $\omega$ Cen should have formed 
before SNIa explosion.

In Figure\,5, the solid line illustrates the MDF given by model~A1, 
while the dashed one gives the MDF obtained by NFM96.
We can say that the modified outflow model
of the bimodal IMF can roughly reproduce the observation
if we focus on the key features of the empirical MDF such as 
the well-defined peak and the metal-rich tail. 
However, discrepancies exist between model~A1 and
the observations. 
At [Ca/H]$>-1.2$, model~A1 predicts more stars than the observations.
A part of this discrepancy might come from the
observational effect of under-sampling as we have
mentioned in section~2. We also note
that the theoretical MDF is inconsistent with
the observations at [Ca/H] $< -1.4$, and hence,
the $\chi^2(=23.2)$ is
slightly worse than the best fit outflow model in section~3.1.

\subsubsection{Model involving infall}
The main weakness of model~A1 is the poor fit of the left
wing of the distribution (Figure~5). 
The observed distribution is
between A1 model and the infall model, 
which is much too wide (Figure~3).
This suggests that a model dominated by a more limited amount of
infall at the beginning, and later on by the modified
outflow model could give better fit. After all, a GC in formation
is expected to benefit from infall; and outflow, even if it can partially
overlap, may need several SNII explosions to become dominant.
Since Cayrel (1986) suggested that star formation with a full 
mass spectrum occurs at a shock between the collapsing gas and 
the SNII-driven wind, we modify the previous model and construct 
an outflow model involving infall. 
Another possible cause of the poor fit of model~A1 
is our assumption that
the mass range of newly formed stars changes abruptly
before and after the first SNII explosion.  
We neglect the possibility of gradually changing 
lower mass cutoff in our model discussed below. 

A chemo-dynamical simulation (Hensler \& Burkert 1990) 
to study the ISM evolution in an open one-zone model 
showed that the gas outflow dominates most of the time while 
the inflow occurs during the phase of small internal pressure.  
Although the inflow discretely happens in their model, 
we assume continuous infall of gas for simplicity. 
The gas evolution is given as below:
\begin{equation}
\frac{dM_{\rm g}(t)}{dt}= -C(t)+E(t)+A(t)-B(t).
\end{equation} 
Other model prescriptions are the same as those of 
the previous outflow model with the bimodal IMF, 
therefore the primordial composition of collapsing gas is assumed. 
The number of combination of parameters becomes 
quite large in this case. Therefore, we first carry out a 
rough surveying calculation to find parameter area 
resulting in $\chi^2$-values smaller than $40$.  
The model grids are the following: 
\begin{eqnarray*}
	\begin{array}{ll}
\tau_{\rm sf}  = 0.001, \, 
\left[ 10^{-2.0+0.3i}~ (i=0,...,5) \right],\,
3~~ {\rm Gyr}, \\

\tau_{\rm of}= 0.001,\, \left[10^{-2.0+0.3i}~
 (i=0,...,5) \right],\,3 ~~{\rm Gyr},\\

\tau_{\rm in}= 0.001,\, \left[10^{-2.0+0.3i}~
 (i=0,...,5) \right],\,3 ~~{\rm Gyr}, \\

x  = 1.0, 1.3, 1.5, 2.0, 2.3, \\
T = 10^{-2.0+i \Delta T}~(i=0,...,29) ~~{\rm Gyr},\\
{\rm [Ca/H]}_0= -1.8,-1.6,-1.5, -1.4, 
	\end{array}
\end{eqnarray*}
\vspace{0.05cm} \\
where 
$\Delta T=7.25 \cdot 10^{-2}$ Gyr.
Next, we simulate models by adopting the following parameters 
by considering results of the previous survey calculation: 
\begin{eqnarray*}
	\begin{array}{llr}
\tau_{\rm sf} = 0.05, 0.07, 0.1, 0.15&{\rm Gyr},\\
\tau_{\rm of}= 0.15, 0.20,0.25&{\rm Gyr},\\
\tau_{\rm in} = 0.02, 0.03,0.04,0.05,0.06&{\rm Gyr},\\
x = 1.8, 1.9, 2.0, 2.1, \\
T = 10.^{-2.0+i \Delta T}~(i=0,...29)&{\rm Gyr},\\
{\rm [Ca/H]}_0 = -\infty, -1.6,-1.5, 
	\end{array}
\end{eqnarray*}
where $\Delta T=7.25 \cdot 10^{-2}$ Gyr. 

In Figure~5, the dotted line shows the best-fit ($\chi^2 = 20.1$) 
model (model~B), of which parameters and skewness are given in Table~1.  
Degrees of freedom of the $\chi^2$ are seven.  
Evolution of $E_{\rm th}/\Omega_{\rm g}$ of model~B
is illustrated in Figure~4. 
The epoch of  $E_{\rm th}(t)=\Omega_{\rm g}(t)$ 
is realized at $T_{\rm w}=0.40$\,Gyr, which is 
well after the epoch ($T=0.31$\,Gyr) of establishing the best-fit MDF.  
Thus model~B does not violate the $T < T_{\rm w}$ criterion.    
Both the left and right wings of the MDF of model~B are  
fitted  to the empirical MDF better than those of model A1. 
However, the peak of the MDF of model B is not sharp enough and 
more stars of [Ca/H] $< -1.4$ are formed than the observation.  
The latter discrepancy is similar to that of the previous infall model 
in section\,3.2. Again, this implies a smaller amount of infall. 
Thus we attempt to rectify the discrepancy by 
stopping the infall.  

Theis, Burkert, and Hensler (1992) reported that 
the pressure of hot gas heated by the supernovae drives 
gas outflow and consequently the collapse of gas is stopped  
when ram pressure is neglected in their chemo-dynamical model 
for massive ($10^{11} - 10^{12} M_\odot$) spheroidal galaxies.  
Since they showed this as an extreme case 
to test influences of the ram pressure in the model,   
the situation might not reflect a realistic one. 
However, they suggested that dwarf galaxies might have a strong 
galactic mass loss even if the ram pressure is effective because of 
a shallower potential well. Since the mass (several $\times 10^6 M_\odot$) 
of $\omega$ Cen is smaller than those of dwarf 
galaxies ($10^7 - 10^9 M_\odot$),
the gas inflow will be terminated by high internal pressure 
by supernovae heating. 
Although an epoch ($T_{\rm SN}$) of termination of the infall 
should be treated as a parameter, 
we simply set $T_{\rm SN}$ at an epoch of peak of SNII rate, since 
$T_{\rm SN}$ should be related to supernovae explosions. 
The following equation gives the gas evolution: 
\begin{equation}
\frac{dM_{\rm g}(t)}{dt}= \left\{
\begin{array}{ll}
-C(t)+E(t)+A(t)-B(t) & (t<T_{\rm SN})\\
-C(t)+E(t)-B(t) & (t \ge T_{\rm SN}). \\
\end{array} \right .
	\label{eqn:fg3}
\end{equation}
Other model prescriptions are the same as before. 
Similarly, we perform a rough surveying calculation, and then 
simulate models with the fine grids of parameters in the 
range of selected values. Model grids of the rough survey calculation are the 
same as those of the previous model. We choose the 
following model grids to find the best-fit model:  
\begin{eqnarray*}
	\begin{array}{lr}
\tau_{\rm sf} =  0.05, 0.06,0.07,0.08,0.09 & {\rm Gyr},\\
\tau_{\rm of}=  0.15, 0.20,0.25 & {\rm Gyr},\\
\tau_{\rm in}=  0.05,0.06,0.07,0.08,0.09 & {\rm Gyr}, \\
x = 1.9, 2.0, 2.1, \\
\vspace{0.2cm}
T = 10^{-2.0+i \Delta T}~(i=0,...,29) & {\rm Gyr},\\
{\rm [Ca/H]}_0= -1.6,-1.5, -1.4,\\  
	\end{array}
\end{eqnarray*}
\vspace{0.1cm}\\
where $\Delta T=7.25 \cdot 10^{-2}$ Gyr. 

In Figure~6, the solid line shows 
the best-fit model (model~C) with $\chi^2=4.5$,  
the degrees of freedom of the $\chi^2$-test become eight. 
Model~C fits to the observed MDF very well, as a 
quite good $\chi^2$-value shows. 
The parameters of model~C are summarised in Table~1 together with 
the $\chi^2$-value and the skewness. 
The epoch of the termination of the inflow, i.e., 
the peak epoch of SNII rate, is realized at $T_{\rm SN}=0.08$~Gyr. 
Model~C is characterised by small $\tau_{\rm in}$ and $\tau_{\rm sf}$ 
($\tau_{\rm in}=\tau_{\rm sf}=0.07$~Gyr), relatively large  
$\tau_{\rm of}$ ($=0.2$\~Gyr) and a steep IMF ($x=1.9$). 
The parameters obtained are roughly the same as those of model~B. 
The evolution of  $E_{\rm th}/\Omega_{\rm g}$ is illustrated 
in Figure~4. 
The epoch of $E_{\rm th}(t)=\Omega_{\rm g}(t)$ is 
$T_{\rm w}=0.31$~Gyr, while the resulting duration of star formation 
is $T=0.28~$Gyr. These satisfy the criterion discussed previously. 
We also note that $T$ is much shorter than a typical lifetime 
($\sim 1-2$~Gyr) of SNIa and is consistent with the argument in section\,2.
Thus, model C gives far better fit than any other models 
studied in this paper. 

\section{Discussion}

\subsection{Abundance gradient}

In Figure~7a, we plot an abundance distribution on the projected surface
of $\omega$ Cen, where the coordinate corresponds to
that of the photographic plate of the ROA catalogue (Woolley 1966) and
the abundances are taken from NFM96.
Metal-poor populations are shown by open squares
([Ca/H]$<-1.5$) and triangles ($-1.5 \le {\rm [Ca/H]} < -1.25$),
while metal-rich populations are given by
filled diamonds ($-1.25 \le {\rm [Ca/H]} < -1.0$)
and triangles ($-1.0 \le$ [Ca/H]).
There is no obvious difference between distributions of
metal-poor and metal-rich stars.
Figure\,7b represents a radial distribution of [Ca/H].
The abundance and the positional data are taken
from NFM96 and Mayor et al. (1997), respectively.
Omega Cen is a flattened GC and the ellipticity changes as a function of
distance from the centre. Although we could have plotted 
an ellipsoidal radius,
we do not attempt this in order to avoid confusion by projection effects.
In this diagram, there is no clear abundance gradient.
However, we notice that metal-rich ([Ca/H]$>-0.5$)
stars do not exist at the outer part
of $\omega$ Cen and that most of stars are located near 
[Ca/H] $\simeq -1.4$ independent of the radius.
To see the abundance distribution at a particular radius, we divide
the samples into four bins and calculate the corresponding MDFs.
We limit the MDFs for [Ca/H]$<-0.5$.
In Figure~7c, we illustrate the MDF in each bin, which is
weighted by the number of stars contained in the corresponding bin.
The peaks appear almost at the same [Ca/H] except for
the outermost bin where MDF has an irregular shape due to
a small number of stars.
There are second peaks for
the MDFs of the innermost two bins
at [Ca/H]$\simeq-0.9$. We do not pay attention to
these peaks because of the concern about the sampling effects.
We find that
contribution from metal-rich stars declines from the inner bin
to the outer.

Now we consider why metal-rich stars decrease toward the outer region.
As we have argued before, the gas outflow must occur during star formation
of $\omega$ Cen.
The potential well of the outer region should be shallower than
that of the inner and the gas in the outer region is more easily
expelled.
Hence duration of star formation in the outer
is expected to be shorter than in the inner,
resulting in less progress of chemical evolution in the outer region.

Gas outflow which starts from the outer part of a system
will also occur in elliptical galaxies, since GCs and ellipticals
are suggested as one parameter family (Yoshii \& Arimoto 1987).
Spectroscopic measurements of Mg$_2$ index (e.g., Carollo, \&
Danziger 1994) of elliptical galaxies
show that the colour gradients come from metallicity changes.
Franx \& Illingworth (1990) found that colour gradients, or
metallicity gradients in elliptical galaxies are related with
local escape velocity and concluded that
the origin of the gradients is the gas outflow because
alternative models such as gas inflow and mergers
should not result in
the correlation between the local escape velocity and the metallicity.
Martinelli, Matteucci, \& Colafrancesco (1998)
investigated whether the metallicity gradients
in elliptical galaxies can be explained by assuming an earlier epoch of
galactic wind at outer region where the potential well is shallower
than the inner region.
If ellipticals have experienced
the gas outflow, their MDFs are expected to have a metal-rich tail.

\subsection{Enrichment in GCs}

The self-enrichment can well explain the metallicity
dispersion of $\omega$ Cen.
Based on their numerical simulation for SNRs in a cloud,
Morgan \& Lake (1989) inferred that
$\omega$ Cen is massive enough to survive against energy injection from
generations of SNII.
A deep gravitational potential could be a reason
why only $\omega$ Cen clearly shows evidence of
gas recycling. In proto-clouds of other smaller GCs, the gas would be
expelled
immediately after the first SNII explosion
because of their shallower potential well.
As we have already discussed, formation of low mass stars
should follow that of high mass stars.
In less massive systems, gas recycling and
star formation with a full spectrum of stellar mass
will be terminated much earlier than in $\omega$ Cen.
If this is the case,
long-lived stars in every GC should be formed
in already enriched gas and their composition will be
almost the same. This implication is consistent with
observed cutoff of lower metallicity of the Galactic GC system and
homogeneous composition for each Galactic GC (e.g. Harris 1991).
The lower metallicity cutoff ([Fe/H]$\la -2$)
seems nearly the same for globular cluster systems (GCSs)
in extragalactic systems as well. For example,
Figure~8 of Harris (1991) shows that
the cutoff is located at [Fe/H]$\la -2$, which is common
for a giant elliptical NGC\,5128 and dwarf ellipticals.
The {\it Hubble Space Telescope} observation (Kundu \& Whitmore 1998)
for the GCS of S0 galaxy NGC\,3115 also shows such a
lower metallicity cutoff, which is estimated [Fe/H]$\la -2$
by using their conversion equation from colour to the iron abundance.
If long-lived stars in the GCs were born in enriched gas, these phenomena
can be naturally explained.
Hence, the idea of self-enrichment in GCs presented by Cayrel (1986)
is favourable. Needless to say, if the metallicity of proto-cloud is 
already higher than [Fe/H]$\simeq -2$, owing to chemical enrichment
by halo field stars, the abundance increase via self-enrichment is
negligible and all stars in GCs should have the same abundances as
their mother clouds.

There is a discrepancy in the time span of star formation 
between the best-fit model (model~C) and 
the work presented by Hughes \& Wallerstein (1998).  
Based on accurate Str\"omgren photometry of stars at the 
main-sequence of $\omega$ Cen,
they plotted a colour-magnitude diagram with isochrones and derived 
the age difference ($\sim 4$\,Gyr). 
On the other hand, model~C has a much shorter span of 0.28~Gyr. 
Since they used the so-called isochrone-fitting method to estimate 
a difference in age, the resulting age span of course 
depends on stellar evolutionary tracks adopted. 
For example, adopting a similar method, 
Noble et al. (1991) concluded that a colour distribution 
in the main sequence of $\omega$ Cen is consistent with the 
metallicity distribution in more evolved stars such as subgiants, 
giants, and horizontal branch stars. 
In addition to this, such a long time span of star formation over a 
lifetime of SNIa ($1-2$~Gyr) is inconsistent with the 
empirical abundance patterns (Norris \& Da Costa 1995) which 
are $\alpha$-enhanced, even in the metal-rich tail.  
Thus the result of model~C can explain 
the chemical properties observed in $\omega$ Cen    
better than that of Hughes \& Wallerstein (1998).  

\section{Conclusion}
We have investigated the origin of the metallicity dispersion 
observed in $\omega$ Cen and explored 
whether the self-enrichment scenario can explain the dispersion. 
Models are confronted with the observed MDFs by NFM96 and 
SK96. The empirical MDF is characterised by a rapid increase 
from lower metallicity to the peak metallicity
and a gradual decrease towards higher metallicity, i.e. a metal-rich tail.
We construct five models. The first three are the closed-box, the outflow, 
and the infall models, which all result in inconsistent MDFs with the observation.    
Next, the outflow models with the bimodal IMF are constructed, 
where long-lived stars are assumed to form after the first SNII explosions. 
This model gives a better fit to the empirical MDFs than any other
first three models, but the fit is not yet perfect.
Finally, we modified the previous model of the bimodal IMF and 
obtained the best-fit model. 
In this model we assume a gas infall at the beginning 
and the bimodal IMF. 
A view of chemical enrichment obtained by the best-fit model 
is as follows: a rapid gas infall 
occurs at the beginning and only massive stars 
form. After the first SNII explosions, 
star formation of a full spectrum range in mass occurs and 
chemical enrichment progresses. 
The gas infall is terminated 
due to SNII heating, which leads to a gas outflow. 
Because of gas consumption and the outflow, 
star formation is ceased. The resulting duration of star formation 
is 0.28~Gyr and is shorter than the typical 
lifetime ($\sim 1-2$~Gyr) of SNIa. 
This is consistent with the observation which showed 
super-solar ratios of [$\alpha$/Fe] 
of stars in $\omega$ Cen (Norris \& Da Costa 1995). 
Thus we conclude that the self-enrichment is 
the origin of the abundance dispersions in $\omega$ Cen.  

Our study shows that the view of the formation and evolution of $\omega$ Cen 
agrees with that of GCs 
predicted by Cayrel (1986): the gas accretes to the centre of a proto-cloud, 
where only massive stars form at first. Subsequently star formation
with a full mass spectrum occurs at a shocked region between the
collapsing gas and the wind driven by supernovae. 
The gas is expelled due to a strong energy
injection by SNII. This leads to termination of star formation. 
This scenario can naturally explain the universal low-metallicity cutoff of
GCSs observed at [Fe/H]$\la -2$ in elliptical galaxies.
All GCs should survive at least one SNII, which enriches the 
surrounding gas. 
Long-lived stars formed after the SNII explosion, and hence 
always contained metals. 
Therefore, $\omega$ Cen may be an analogue of small spheroidal galaxies and 
provides the simplest and best opportunity currently
available to investigate how such a system chemically enriches itself.
We expect that similar studies of the MDFs 
will reveal chemical enrichment 
and star formation histories of dwarf galaxies in the Local Group.

\begin{acknowledgements}
The authors are both grateful and indebted to the referee, R. Cayrel, 
for his helpful remarks and suggestions 
on the early version of their models and the paper. 
C.I. thanks T. Kodama for fruitful discussion and 
the Japan Society for Promotion of 
Science for a financial support. 
This work was financially supported in part 
by a Grant-in-Aid for Scientific Research (No. 11640230) 
by the Japanese Ministry of Education, Culture, Sports
and Science. 
\end{acknowledgements}


\clearpage
\begin{table*}
\begin{center}
\begin{tabular}{crcccccccccr}
\hline
\hline
Model & $\chi^2$ &[Ca/H]$_0$ & $x$ & $\tau_{\rm sf}$ & $\tau_{\rm of}$ & $\tau_{\rm in}$ 
& $T$ & [Ca/H] & $f_{\rm g}$ & $T_{\rm W}$ & skew \\
\hline
A1 & 23.2 & -1.40 &  1.3 & 0.160 & 0.010  & $-$ & 0.030 & -0.56 & 0.03 & 0.030 &0.89\\
A2 & 25.8 & -1.40 &  1.9 & 0.025 & 0.025 & $-$ & 0.060 & -0.53 & 0.01 & 0.042& 0.95 \\
B & 15.8 & -1.60 &1.9 & 0.100 & 0.200 & 0.030 & 0.310 & -0.79 & 0.02 & 0.400 & 0.38 \\
C & 4.5 & -1.60 & 1.9 & 0.070 & 0.200 & 0.070 & 0.280 & -0.65 & 0.006 &  0.310 & 0.58 \\
\hline
\end{tabular}
\end{center}
\caption{
Parameters of models A1, A2, B, and C described in subsection 3.3. 
Column\,(1): model identification,  
Column\,(2): fitting goodness, Column\,(3): initial calcium abundance, 
Column\,(4): IMF power, Column\,(5): time scale of star formation in Gyr, 
Column\,(6): time scale of outflow, 
Column\,(7): time scale of inflow, 
Column\,(8): duration of star formation,  
Columns\,(9) \& (10): [Ca/H] and gas fraction at $T$,  
Column\,(11): epoch of $E_{\rm th} = \Omega_{\rm g}$ in Gyr, 
Column\,(12): skewness of each model. The skewness 
of the empirical MDF of NFM96 is $0.73$
}


\end{table*}

\clearpage

\figcaption[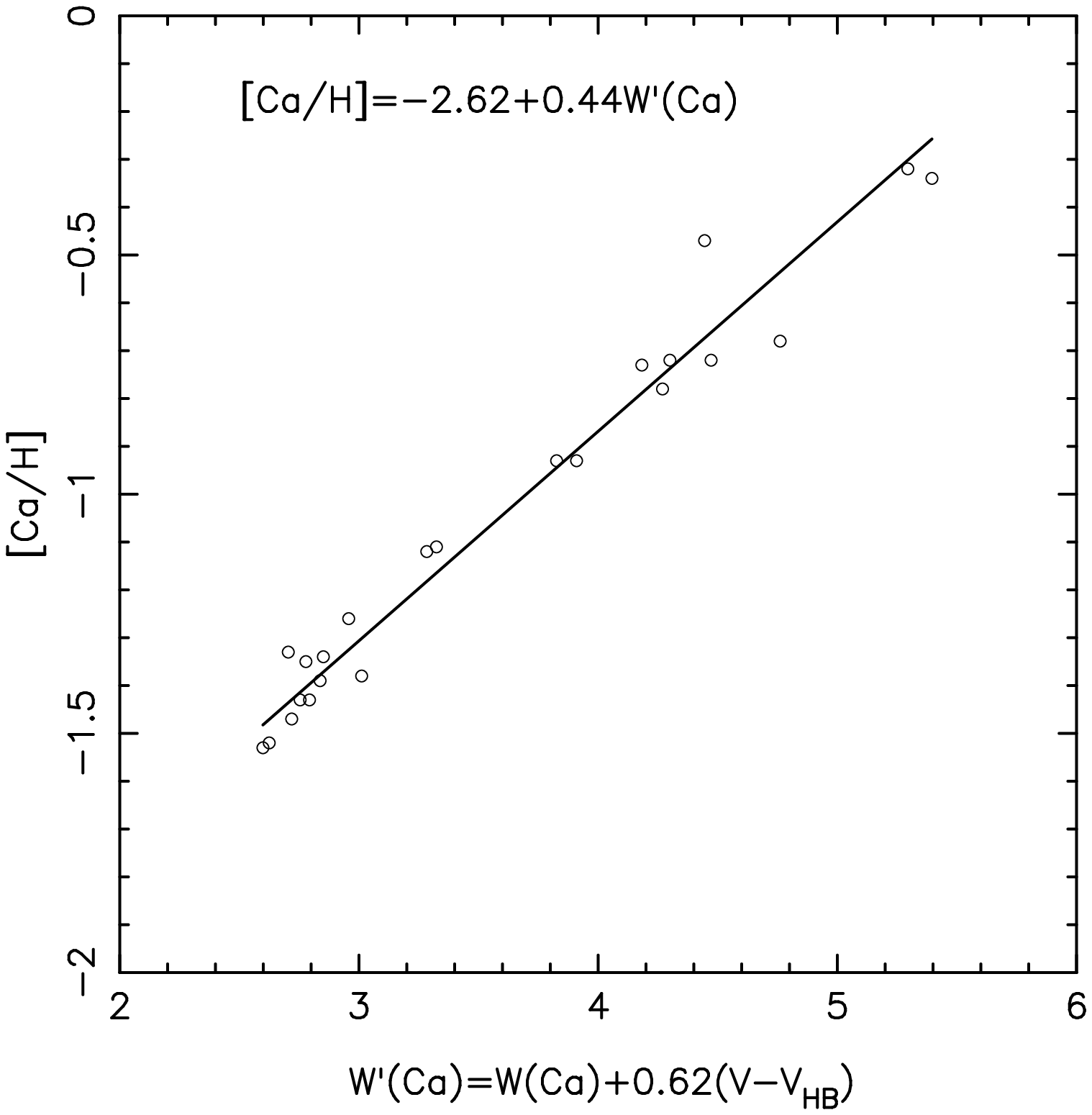]{Calibration of [Ca/H] as a function of the reduced 
calcium equivalent width W'(Ca) of the BG sample by SK96 in units of 
\AA. The abundance data by high dispersion spectroscopy were taken 
from Norris \& Da Costa (1995).\label{fig1}}
	
\figcaption[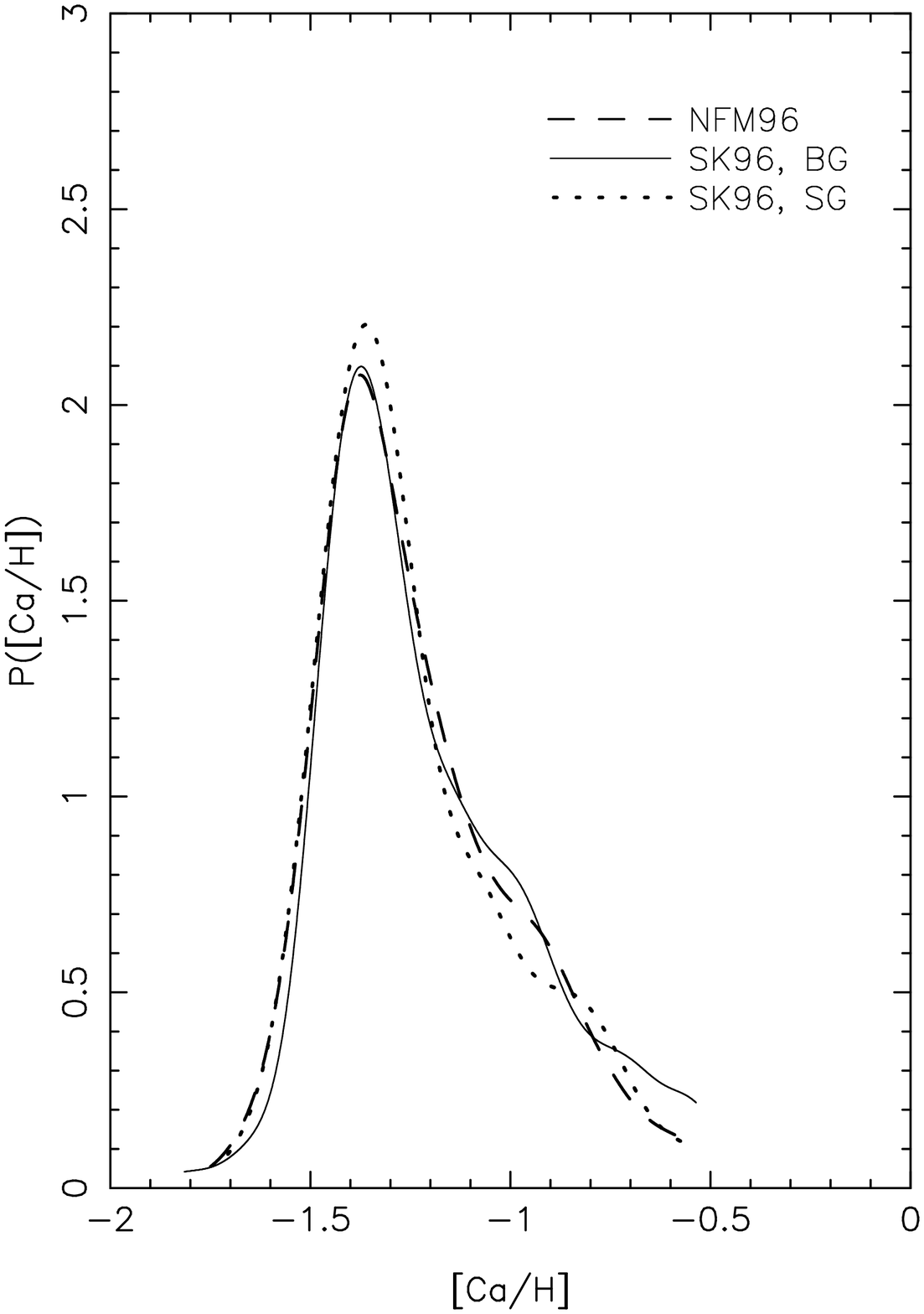]{
Observed metallicity distribution functions of [Ca/H]. 
Dashed line represents the sample observed by NFM96. 
Solid and dotted lines give the BG and the SG samples (SK96), 
respectively. 
\label{fig2}}

\figcaption[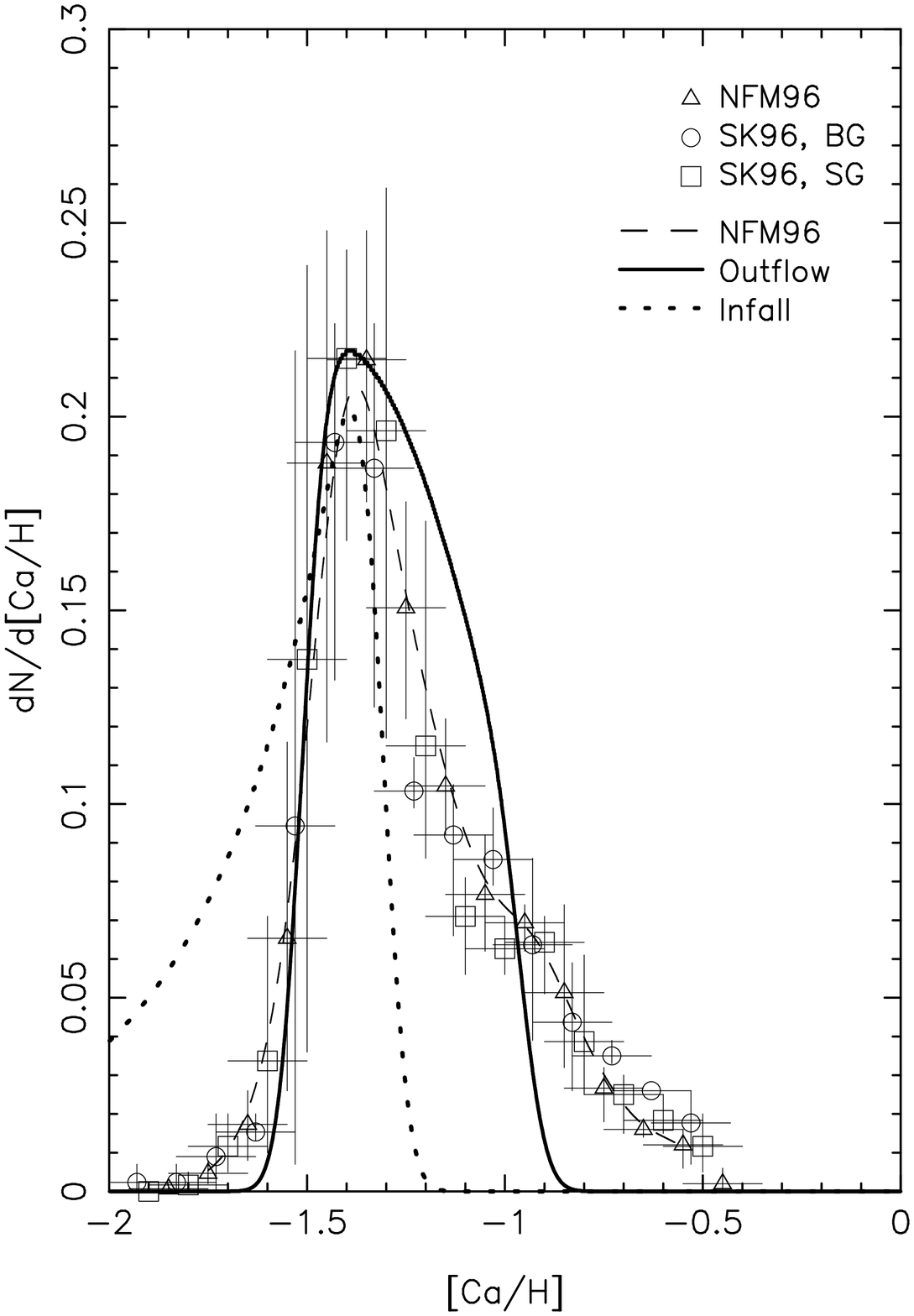]{
A comparison of the empirical and the theoretical MDFs. The 
solid line represents the best fit MDF of the outflow model described in 
section 3.1,  and the dotted line gives that of the infall model 
in section 3.2. The dashed line illustrates 
the empirical MDF obtained by NFM96's data.   
Symbols indicate each sample as following: ($triangles$) NFM96; 
($circles$) the BG of SK96; ($squares$) the SG of SK96.  
The vertical axis gives a number normalised by the sum of each sample.
\label{fig3}}

\figcaption[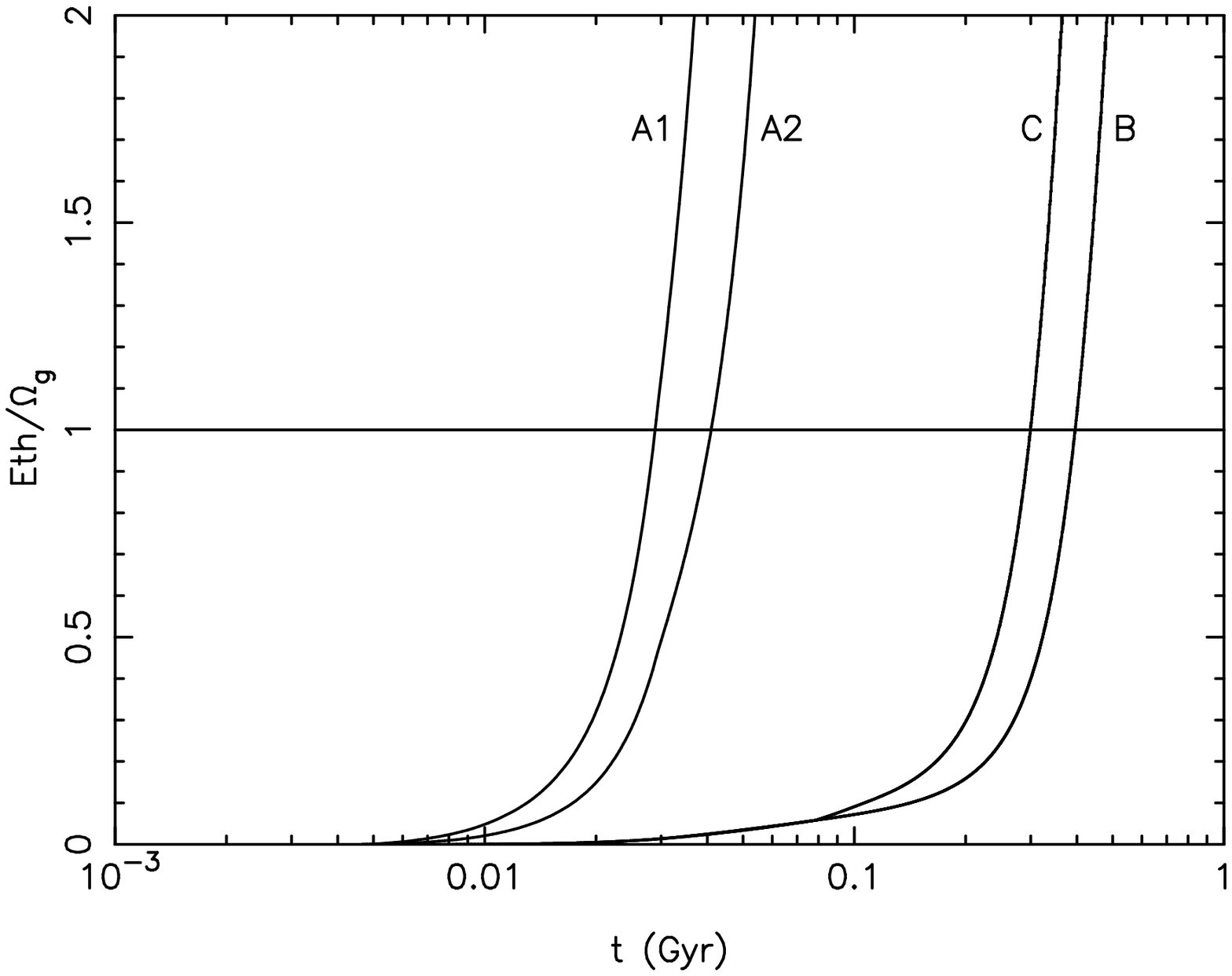]
{Evolution of the ratio between the cumulative thermal energy 
$E_{\rm th}(t)$ from SNII and the gas binding energy $\Omega_{\rm g}(t)$ for 
models A1, A2, B, and C. The best fit MDFs of these models 
are realized at 0.03, 0.06, 0.31, and 0.28~Gyr, respectively. \label{fig4}}

\figcaption[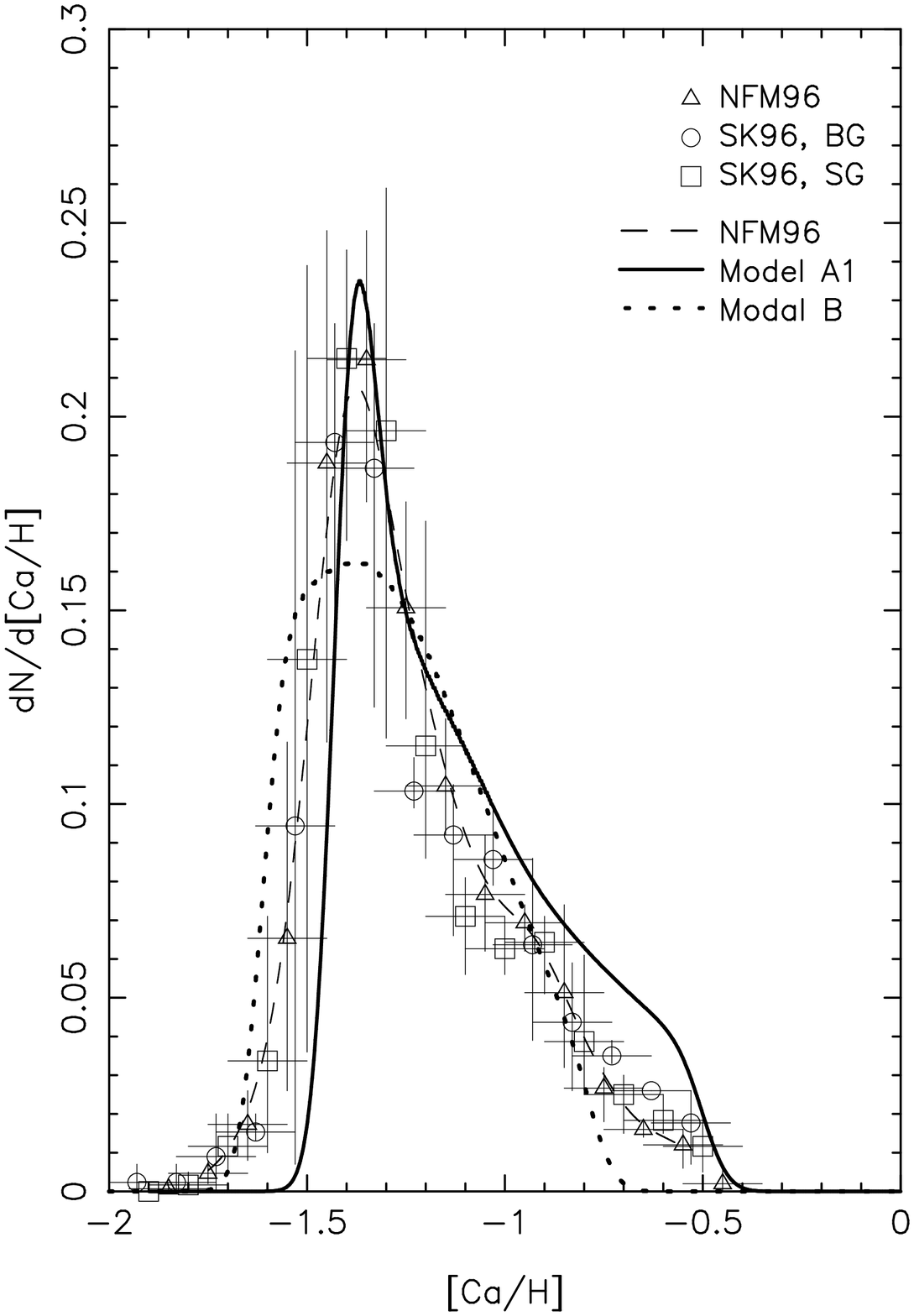]{
Theoretical MDFs of models A1 (solid line) 
and B (dotted line). The dashed line gives the observed MDF of NFM96 sample. 
Symbols have the same meanings as those of Figure~3. \label{fig5}}

\figcaption[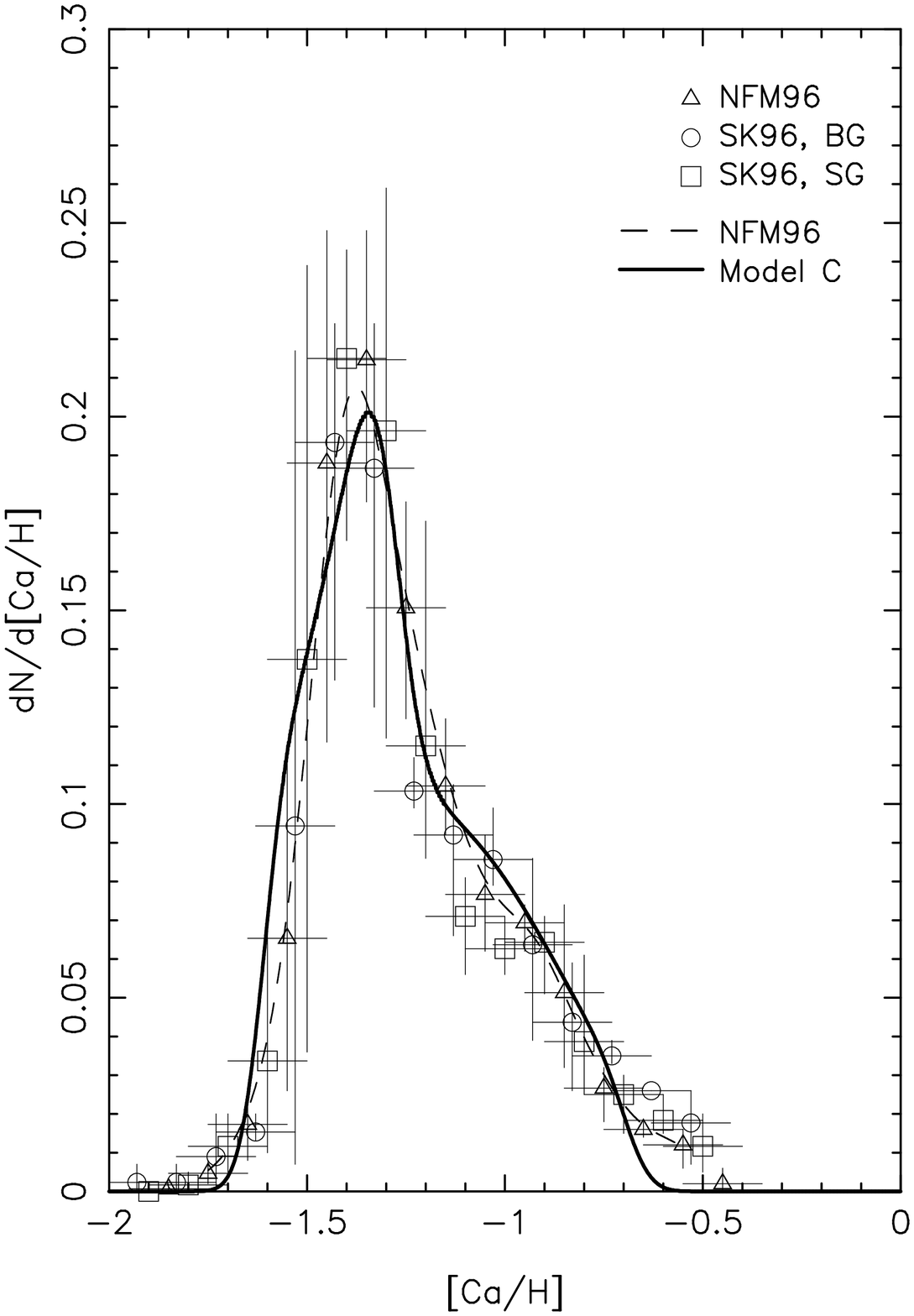]
{Theoretical MDF of model C (solid line) and 
the observed MDF (dashed line) of NFM96 sample. 
Symbols have the same meanings as those of Figure~3. 
\label{fig6}}

\figcaption[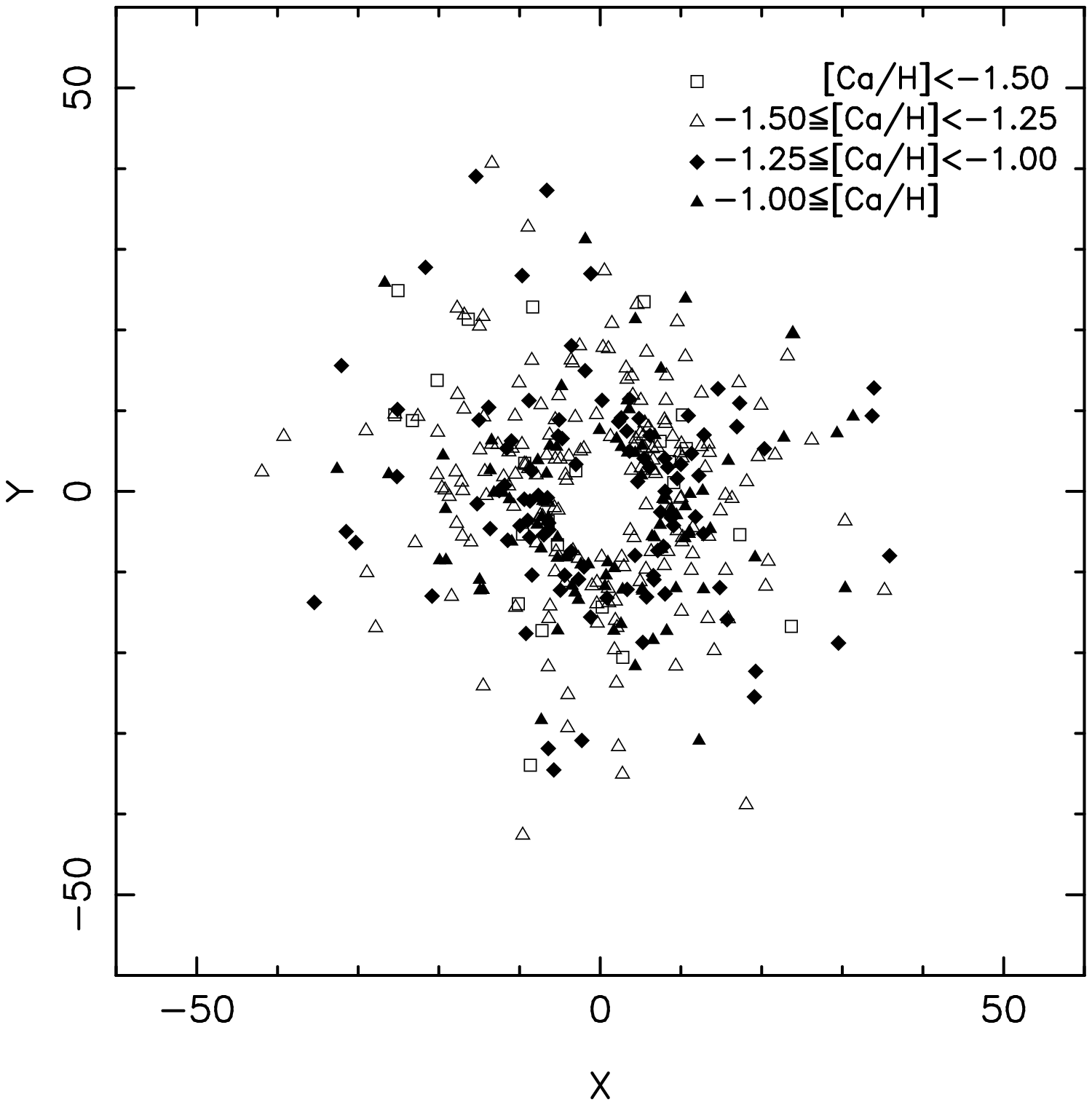,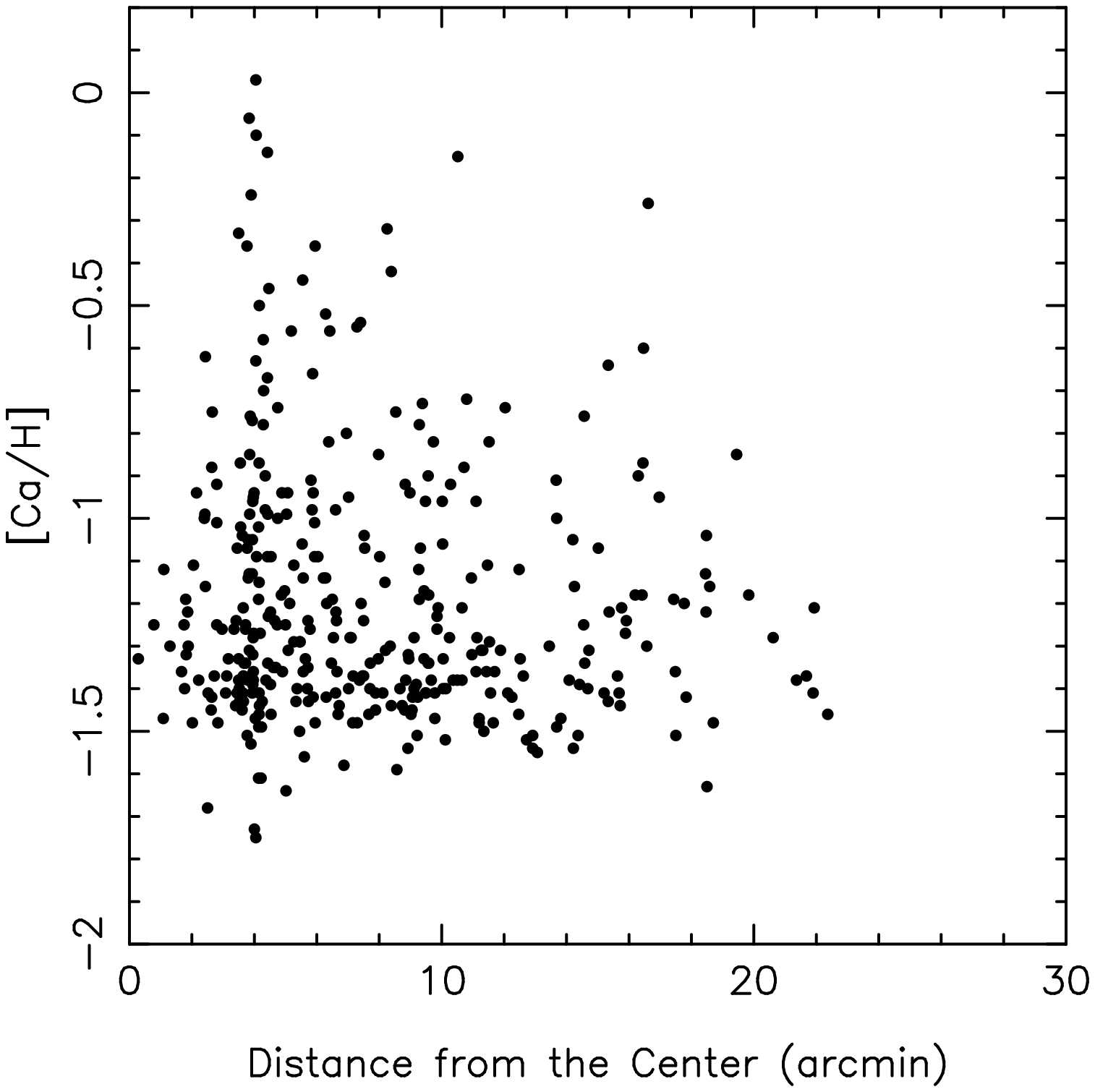,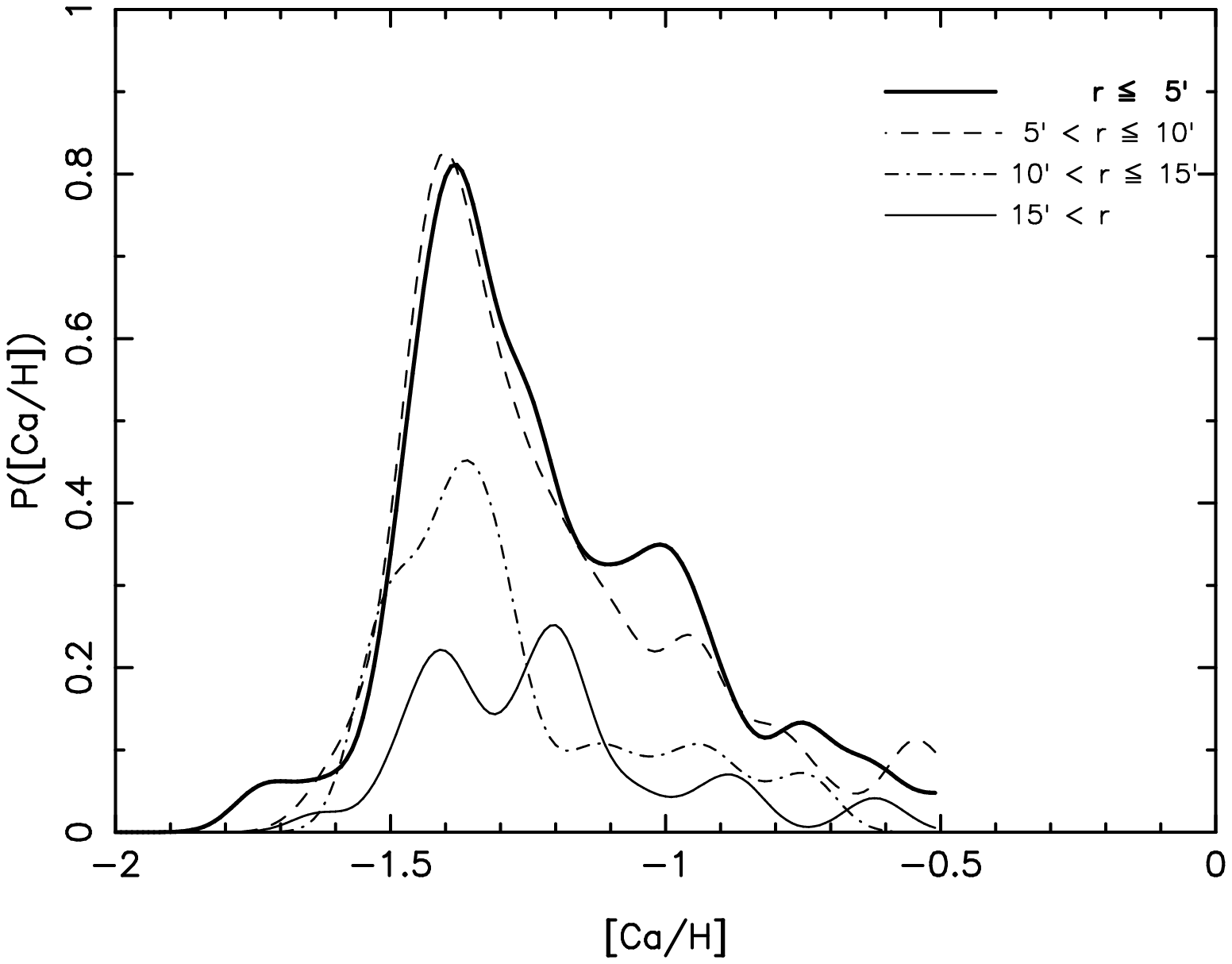]
{(a) A map of calcium abundance. The coordinate corresponds to 
that of the ROA catalogue (Woolley 1966). The abundance data 
is taken from NFM96. 
Each symbol represents different abundance of stars, which is indicated in
the panel.     
(b) Dependence of [Ca/H] on radial distance. The abundance data and 
the distance are taken from NFM96 and Mayor et al (1997), respectively. 
(c) A comparison of MDFs at different distance from the centre. 
They are weighted by a number of stars in each bin. 
Lines are labelled inside the panel with the distance range of 
the bin. \label{fig6}}

\clearpage
\plotone{9014.f1.eps}

\plotone{9014.f2.eps}

\plotone{9014.f3.eps}

\plotone{9014.f4.eps}

\plotone{9014.f5.eps}

\plotone{9014.f6.eps}

\plotone{9014.f7a.eps}

\plotone{9014.f7b.eps}

\plotone{9014.f7c.eps}

\end{document}